\documentstyle[wstwocl,epsf]{article}
\begin{document}

\bibliographystyle{unsrt}    

\newcommand{\st}{\scriptstyle}
\newcommand{\sst}{\scriptscriptstyle}
\newcommand{\mco}{\multicolumn}
\newcommand{\epp}{\epsilon^{\prime}}
\newcommand{\vep}{\varepsilon}
\newcommand{\ra}{\rightarrow}
\newcommand{\ppg}{\pi^+\pi^-\gamma}
\newcommand{\vp}{{\bf p}}
\newcommand{\ko}{K^0}
\newcommand{\kb}{\bar{K^0}}
\newcommand{\al}{\alpha}
\newcommand{\ab}{\bar{\alpha}}
\def\be{\begin{equation}}
\def\ee{\end{equation}}
\def\bea{\begin{eqnarray}}
\def\eea{\end{eqnarray}}
\def\CPbar{\hbox{{\rm CP}\hskip-1.80em{/}}}

\def\ap#1#2#3   {{\em Ann. Phys. (NY)} {\bf#1} (#2) #3}
\def\apj#1#2#3  {{\em Astrophys. J.} {\bf#1} (#2) #3}
\def\apjl#1#2#3 {{\em Astrophys. J. Lett.} {\bf#1} (#2) #3}
\def\app#1#2#3  {{\em Acta. Phys. Pol.} {\bf#1} (#2) #3}
\def\ar#1#2#3   {{\em Ann. Rev. Nucl. Part. Sci.} {\bf#1} (#2) #3}
\def\cpc#1#2#3  {{\em Computer Phys. Comm.} {\bf#1} (#2) #3}
\def\err#1#2#3  {{\it Erratum} {\bf#1} (#2) #3}
\def\ib#1#2#3   {{\it ibid.} {\bf#1} (#2) #3}
\def\jmp#1#2#3  {{\em J. Math. Phys.} {\bf#1} (#2) #3}
\def\ijmp#1#2#3 {{\em Int. J. Mod. Phys.} {\bf#1} (#2) #3}
\def\jetp#1#2#3 {{\em JETP Lett.} {\bf#1} (#2) #3}
\def\jpg#1#2#3  {{\em J. Phys. G.} {\bf#1} (#2) #3}
\def\mpl#1#2#3  {{\em Mod. Phys. Lett.} {\bf#1} (#2) #3}
\def\nat#1#2#3  {{\em Nature (London)} {\bf#1} (#2) #3}
\def\nc#1#2#3   {{\em Nuovo Cim.} {\bf#1} (#2) #3}
\def\nim#1#2#3  {{\em Nucl. Instr. Meth.} {\bf#1} (#2) #3}
\def\np#1#2#3   {{\em Nucl. Phys.} {\bf#1} (#2) #3}
\def\pcps#1#2#3 {{\em Proc. Cam. Phil. Soc.} {\bf#1} (#2) #3}
\def\pl#1#2#3   {{\em Phys. Lett.} {\bf#1} (#2) #3}
\def\prep#1#2#3 {{\em Phys. Rep.} {\bf#1} (#2) #3}
\def\prev#1#2#3 {{\em Phys. Rev.} {\bf#1} (#2) #3}
\def\prl#1#2#3  {{\em Phys. Rev. Lett.} {\bf#1} (#2) #3}
\def\prs#1#2#3  {{\em Proc. Roy. Soc.} {\bf#1} (#2) #3}
\def\ptp#1#2#3  {{\em Prog. Th. Phys.} {\bf#1} (#2) #3}
\def\ps#1#2#3   {{\em Physica Scripta} {\bf#1} (#2) #3}
\def\rmp#1#2#3  {{\em Rev. Mod. Phys.} {\bf#1} (#2) #3}
\def\rpp#1#2#3  {{\em Rep. Prog. Phys.} {\bf#1} (#2) #3}
\def\sjnp#1#2#3 {{\em Sov. J. Nucl. Phys.} {\bf#1} (#2) #3}
\def\spj#1#2#3  {{\em Sov. Phys. JEPT} {\bf#1} (#2) #3}
\def\spu#1#2#3  {{\em Sov. Phys.-Usp.} {\bf#1} (#2) #3}
\def\zp#1#2#3   {{\em Zeit. Phys.} {\bf#1} (#2) #3}
\renewcommand{\thefootnote}{\fnsymbol{footnote}}

\setcounter{secnumdepth}{2} 


\title{\vspace{2.em}
TOP QUARK PAIR PRODUCTION IN THE THRESHOLD REGION\footnotemark[1]
\vspace{-4.5em}
\begin{flushright}
\large\bf
TTP95--46\footnotemark[2]\\
November 1995\\
hep-ph/9512331
\end{flushright}
\vspace{1.5em}
}

\firstauthors{Marek Je\.zabek}

\firstaddress{Institute of Nuclear Physics, ul. Kawiory 26a,
PL-30055 Cracow, Poland}

\twocolumn[\maketitle\abstracts{
Recent results on production and decays of polarized top quarks
are reviewed.
Top quark pair production in $e^+e^-$ annihilation is considered
near energy threshold.
For longitudinally
polarized electrons the produced top quarks and antiquarks
are highly polarized.
Dynamical effects originating from strong interactions
and Higgs boson exchange in the
$t-\bar t$ system can be calculated using the Green function method.
Energy-angular distributions of leptons in semileptonic decays
are sensitive to the polarization of the decaying top
quark and to the Lorentz structure of the weak charged current.
}]

\section{Introduction}
\footnotetext[1]{\hspace{.5em}
	Invited talk presented on International Europhysics
	Conference on High Energy Physics, Brussels, 27.7.--2.8.1995,
	to appear in proceedings.}
\footnotetext[2]{\hspace{.5em}
	The complete paper is also available via anonymous ftp at
	ftp://www-ttp.physik.uni-karlsruhe.de/, or via www at
	http://www-ttp.physik.uni-karlsruhe.de/cgi-bin/preprints/.}
As the heaviest fermion of the Standard Model the top quark
is an exciting new window on very high mass scale physics.
There is no doubt that precise studies of top quark
production and decays will provide us with new information
about the mechanism of electroweak symmetry breaking.
The analysis of polarized top quarks and their decays
has recently attracted considerable
attention\cite{Kuehn3,teupitz,kzfest}.
For non-relativistic top quarks the polarization studies
are free from hadronization ambiguities. This is due to the
short lifetime of the top quark which is shorter than
the formation time of top mesons and toponium resonances.
Therefore top decays interrupt the process of hadronization
at an early stage and practically eliminate associated
non-perturbative effects.

The most efficient and flexible reaction producing
polarized top quarks is pair production in $e^+e^-$
annihilation with longitudinally polarized electron beams.
For $e^+e^-\to t\bar t$ in the threshold region
one can study decays of polarized top quarks under
particularly convenient conditions: large event rates,
well-identified rest frame of the top quark,
and large degree of polarization. At the same time,
thanks to the spectacular success
of the polarization program at SLC \cite{woods},
the longitudinal polarization of the electron beam
will be an obvious option for a future linear collider.

In the present article some recent results\cite{HJKT,HJK,HJKP}
are presented for polarized top
quark pair production in $e^+e^-$ annihilation
near production threshold. The Green
function method \cite{FK,SP,JKT,Sumino1}
has been extended to the case of polarized $t$ and $\bar t$.
It has been shown that for the longitudinally
polarized electron beam longitudinally polarized top quarks
can be produced. The transverse and normal components of
the top quark polarization have been calculated
including effects of $S-P$ interference and
rescattering corrections.
Effects of Higgs particle exchange have been also studied.
Semileptonic decays of polarized top quarks are briefly
discussed. The cleanest spin analysis for the top quarks
can be obtained from their semileptonic decay
channels\cite{teupitz,jk94}.

\section{Pair production near threshold}
The top quark is a short--lived particle. For the top mass
$m_t$ in the range 160--190 GeV its width
$\Gamma_t$ increases with $m_t$ from 1 to 2 GeV.
Thus $\Gamma_t$  by far exceeds
the hyperfine splitting for toponia and
open top hadrons,
the hadronization scale of about 200 MeV,
and even the energy splitting between $1S$ and $2S$ $t\bar t$
resonances. Toponium
resonances including the $1S$ state overlap each other.
As a consequence the cross section for $t\bar t$ pair production
near energy threshold has a rather simple and smooth shape.
It can be shown\cite{FK,SP}  that
for non-relativistic $t$ and $\bar t$
the dominant contribution to the amplitude is given
by the sum of the ladder diagrams
corresponding to chromostatic interactions between top and antitop.
The other contributions are suppressed
by factors of order $\beta^2$ where $\beta$ denotes
the velocity of the top quark in the center-of-mass frame.
The ladder diagram with $n$ exchanges
of gluons
gives the contribution of order $(\alpha_s/\beta)^n$ where
$\alpha_s$ is the strong coupling constant.
In the threshold region $\beta \sim \alpha_s$ and
all the contributions are of the same order.
The sum of the ladder diagrams
can be expressed through the Green function of the $t-\bar t$
system. The effects of the top quark width are
incorporated through the complex energy $E\,+\,i\Gamma_t$,
where
$E=\sqrt{s} - 2m_t$
is the non-relativistic energy of the system.
The idea \cite{FK,SP} to use the Green function
instead of summing over overlapping resonances has been also
applied in numerical calculations of differential cross sections.
Independent approaches have been developed for solving Schr\"odinger
equation in position space \cite{Sumino1} and Lippmann-Schwinger
equation in momentum space \cite{JKT,JT}.
The results of these two methods agree very well.
One of the most important future applications\cite{Miquel} will be
the determination of $m_t$ and $\alpha_s$.
It has been pointed out\cite{Sumino1} that the momentum dependence
of the width for $t-\bar t$ system which arises at order $\alpha_s^2$
may be important in quantitative studies.
However, it has been conjectured\cite{JT}
that order $\alpha_s^2$ rescattering corrections
to the width nearly cancel the negative contributions
originating from phase space suppression.
The remainder can be interpreted as due to time dilatation
for $t$ and $\bar t$ in the center-of-mass frame. Its effect
on the total cross section is quite small and can be neglected.
Recently an elegant proof of this conjecture has been
found\cite{KuM}.

\begin{figure}[tb]
\begin{center}
\epsfxsize=8.0cm
\leavevmode
\epsffile[30 310 535 510]{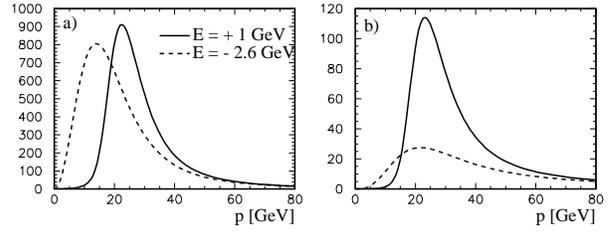}
\vskip-0.2cm
\caption{Top quark momentum -- angular distribution
functions: a) ${\cal D}_{S-S}(p,E)$ and
b) ${\cal D}_{S-P}(p,E)$
for $m_t = 174$ GeV.
\label{fig-brufig1}}
\end{center}
\end{figure}

The Green function method has been generalized
to the case of polarized top quark
pair production in $e^+e^-$ annihilation\cite{HJKT,kzfest}
and in $\gamma\gamma$ collisions\cite{fkk}.
In the non-relativistic approximation,
neglecting contributions of order $\beta^2$
the calculation of polarized cross sections can be
reduced to solving of the Lippmann-Schwinger equations
for the $S$ wave and for the $P$ wave Green functions:
\begin{equation}
G\left( p,E\right) =
G_0(p,E)\> \left[\>
1 +
\int {d^3 q\over(2\pi)^3}\, V\left({\bf p - q} \right)\,
G(q,E) \>  \right]
\end{equation}
\begin{equation}
F\left( p,E\right)
\hskip-1pt = \hskip-1pt
G_0(p,E) \left[ 1 + \hskip-2pt
\int {d^3 q\over(2\pi)^3} \hskip-2pt
{{\bf p\cdot q}\over p^2}
V\left({\bf p - q} \right)
F(q,E)\right]
\end{equation}
where $\bf p$ is top quark momentum, $G_0(p,E)$
denotes the free Green function for the $t-\bar t$ system
\begin{eqnarray}
G_0(p,E) \hskip-5pt
&=& \hskip-5pt
\>=\> {\textstyle \left(\> E - {p^2\over m_t} +
{\rm i}\Gamma_t\> \right)^{-1} }
\end{eqnarray}
and $V\left({\bf p - q} \right)$ denotes the chromostatic
potential in momentum space. In our numerical calculations
a potential\cite{JT} was used given by the two-loop
perturbative formula at large momenta and by Richardson's
potential at intermediate and small ones.
The $S$ wave Green functions $G$
governs the momentum distribution
of the top quark
\begin{eqnarray}
{d\sigma\over dp}\; \sim \;
{\cal D}_{S-S}(p,E) &=& p^2 \left|\, G(p,E)\,\right|^2
\end{eqnarray}
Interference between $S$ and $P$ partial waves\cite{MuS}
results in a non-trivial momentum and angular distribution
which for $e^+e^-$ annihilation and linearly polarized beams
reads\cite{HJKT}:
\begin{eqnarray}
\hskip-2pt
{d\sigma\over dp\, d\Omega} \, \sim \,
{\cal D}_{S-S}(p,E)\>\left[\, 1\, +\, 2\, C_{FB}(\chi) \,
\varphi_R(p,E)\,\cos\vartheta \right]
\end{eqnarray}
where $\vartheta$ is the angle between the electron and the top
quark, and
$\varphi_R(p,E)$ denotes the real
part of the function
\begin{equation}
\varphi(p,E)\, =\,  {
\left( 1- {4\alpha_s\over 3\pi} \right)\> p\, F^*(p,E)
\over
\left( 1- {8\alpha_s\over 3\pi} \right)
m_t\, G^*(p,E) }
\label{eq-phi}
\end{equation}
$C_{FB}$ depends on the electroweak charges of electron and
top quark and the variable
\begin{equation}
\chi = {P_{e^+}-P_{e^-}\over 1 - P_{e^+}P_{e^-}}
\end{equation}
where $P_{{e^+}}$ and $P_{{e^-}}$ denote the linear polarizations
of the beams.
The functions ${\cal D}_{S-S}(p,E)$ and
\begin{eqnarray}
{\cal D}_{S-P}(p,E) &=& {\cal D}_{S-S}(p,E)\> \varphi_R(p,E)
\end{eqnarray}
are shown in Fig.\ref{fig-brufig1} for two energies close to
the threshold.
To describe polarized top quark production in the
threshold region it is convenient to
define the components of its
polarization vector with respect to the
triplet of orthogonal unit vectors: $\hat n_{^\|}$,
$\hat n_{\bot}$ and $\hat n_{_N}$, where
$\hat n_{^\|}$ points in the direction of the $e^-$ beam,
$\hat n_{_N}\sim \vec p_{e^-}\times \vec p_t$  is normal
to the production plane and
$\hat n_{\bot}=\hat n_{_N}\times\hat n_{^\|}$.
Retaining only the terms up to
${\cal O}(\beta)$ one derives\cite{HJKT} the following expressions
for the components of the polarization vector, as functions
of $E$, $p$, $\vartheta$ and $\chi$:
\begin{eqnarray}
\hskip-5pt
{\cal P}_{^\|}(p,E,\vartheta,\chi) &=& C^0_{^\|}(\chi)
+ C^1_{^\|}(\chi)\,\varphi_R(p,E)\,\cos\vartheta
\label{eq-Ppar}\\
{\cal P}_\bot(p,E,\vartheta,\chi) &=&C_\bot(\chi)\,
\varphi_R(p,E)\, \sin\vartheta
\label{eq-Pperp}\\
{\cal P}_N(p,E,\vartheta,\chi) &=&C_N(\chi)\,
\varphi_I(p,E)\, \sin\vartheta
\label{eq-Pnorm}
\end{eqnarray}
where $\varphi_R(p,E)$ and $\varphi_I(p,E)$ denote the real
and imaginary parts of the function $\varphi(p,E)$
defined in eq.(\ref{eq-phi}).
\begin{figure}
\begin{center}
\epsfxsize=8.0cm
\leavevmode
\epsffile[30 310 535 510]{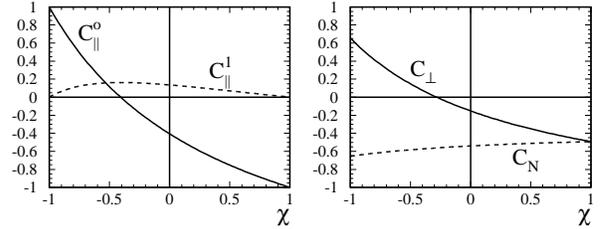}
\vskip-0.2cm
\caption{Coefficient functions: a) $C^0_{^\|}(\chi)$ -- solid line
and $C^1_{^\|}(\chi)$ -- dashed line, b) $C_\bot(\chi)$ -- solid line
and $C_N(\chi)$ -- dashed line.
\label{fig-Cppn}}
\end{center}
\end{figure}
\begin{figure}
\begin{center}
\epsfxsize=8.0cm
\leavevmode
\epsffile[30 310 535 510]{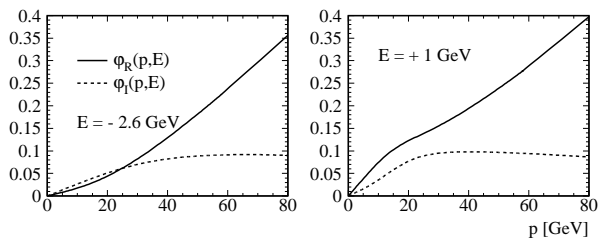}
\vskip-0.2cm
\caption{Momentum dependence of the functions $\varphi_R(p,E)$
(solid lines) and  $\varphi_I(p,E)$ (dashed lines):
a) $E$ = -2.6 GeV, and b) $E$ = 1 GeV.
\label{fig-phiRI}}
\end{center}
\end{figure}
The energy dependence of all the coefficient functions
$C(\chi)$ is very weak and can be neglected.
In Fig.\ref{fig-Cppn}a the coefficient functions $C^0_{^\|}(\chi)$
and $C^1_{^\|}(\chi)$ are shown. It is evident that for maximal
and minimal values of $\chi=\pm1$ the top quark is maximally
polarized along the direction of the incoming electron.
The correction
to the parallel polarization is quite small, so,  to a good
approximation
${\cal P}_{^\|}$ and $C^0_{^\|}(\chi)$ are equal. The shape of
the latter function depends on the electroweak charges of the top
quark. For example if we change the values of $Zt\bar t$ couplings
rotating the $(v_t,a_t)$ vector by $\pm 0.1 rd$ then
$C^0_{^\|}(0.8)= -.91\pm.02$, where the central value denotes the
Standard Model result, $C^0_{^\|}(0)= -.41^{+.10}_{-.09}$,  and
$C^0_{^\|}(-0.8)= .58^{+.07}_{-.08}$. Thus, measuring the top quark
parallel polarization as a function of $\chi$ one can measure $v_t$
and $a_t$.
This demonstrates that polarization studies close to threshold
are very promising indeed. The other components of the top
polarization can be also interesting and the corresponding
coefficient functions are plotted in Fig.\ref{fig-Cppn}b.
Momentum dependence of the functions $\varphi_R(p,E)$
and  $\varphi_I(p,E)$ is shown in Fig.\ref{fig-phiRI}
for two energies in the threshold region.

\section{Semileptonic decays of top quarks}
The energy and angular distributions of the charged leptons and
the neutrinos are sensitive to the polarization of the decaying
heavy quark. Therefore they can be used in determination of
this polarization.
Compact analytic formulae have been recently found\cite{CJ}
for QCD corrections to the energy-angular distributions of
the charged leptons and the neutrinos in semileptonic decays
of the top quark.
In the rest frame of the decaying top quark
the double differential energy-angular distribution
of the charged lepton is the product of the energy
distribution and the angular distribution\cite{JK89}.
QCD corrections
essentially do not spoil this factorization\cite{CJK91}.
Thus the polarization analysing power of
the charged lepton energy-angular distribution
is maximal and hence far superior to other
distributions discussed in the following. In particular
for the neutrino energy-angular distribution factorization
does not hold.
It follows that the charged lepton
angular distribution in $t\to Wb\to b\ell^+\nu$ decays
is significantly
more sensitive towards the polarization of $t$ than the
angular distributions of $W$ and $\nu$.
The neutrino distributions, however, are more sensitive
to deviations from $V-A$ and can be used in testing this
basic assumption about the Lorentz structure
of the charged weak current\cite{jk94}.
The charged lepton is likely to be the less energetic
lepton because its energy spectrum is softer
than that of the neutrino.
For large values of $m_t$ the angular distribution
of the less energetic lepton
is a more efficient analyser of top polarization than the angular
distribution of neutrinos. For $m_t$ in the range 150-200 GeV
it is also better than the direction of $W$ boson. For semileptonic
$t$ decays this observation is not very useful.
However, for the hadronic
$t\to Wb\to b{\bar u}d$ decays the angular distribution of the
less energetic jet can be used\cite{teupitz}.

\setcounter{secnumdepth}{0} 

\section{Acknowledgments}
I would like to thank the members of the Institut f\"ur
Theoretische Teilchenphysik, Universit\"at Karlsruhe
where a large fraction of the work reported in this
article was done. In particular I am very indebted to Hans K\"uhn,
Andrzej Czarnecki,
Robert Harlander, Markus Peter and Thomas Teubner
for common research on top quark production and decays.\\
This work was supported by KBN Grants 2P30225206 and 2P30207607,
and by EEC Contract CIPDCT940016.

\end{document}